\documentstyle[eqsecnum,aps,prb,multicol,epsfig]{revtex}

\begin{document}

\title{Variational Theory of Flux Line Liquids}
\author{A. M. Ettouhami}
\address{Department of Physics, University of Colorado, Boulder, Colorado 80309}
\date{\today}
\maketitle

\begin{abstract}

We formulate a variational (Hartree like) description of flux line liquids which improves on the theory we
developed in an earlier paper [A.M. Ettouhami, Phys. Rev. B {\bf 65}, 134504 (2002)].
We derive, in particular, how the massive term confining the fluctuations of flux lines varies with temperature and 
show that this term vanishes at high enough temperatures where the vortices behave as freely fluctuating elastic 
lines.

\end{abstract}

\pacs{74.20.De, 74.60.Ge}

\begin{multicols}{2}

\section{Introduction}
\label{Introduction}

The study of the properties of flux line solids and liquids in high temperature superconductors (HTSC) has been, during 
the past few years, one of the most active areas of research in the field of superconductivity. 
One theoretical approach that has had a rather strong impact on our present understanding of flux liquids in HTSC is 
the boson mapping
\cite{Nelson,Nelson-Seung,Nelson-LeDoussal,Tauber-Nelson,Benetatos-Marchetti}, 
which is based on the observation \cite{Fisher} that there is 
a formal mapping between the partition function of a three-dimensional system of interacting flux lines, and the 
imaginary-time partition function of quantum bosons in two-dimensions. One of the main results of this approach, which 
has in fact been reproduced using other methods\cite{Radzihovsky-Frey}, is that the structure factor 
\begin{eqnarray}
S({\bf r},z)=\langle\hat\rho({\bf r},z)\hat\rho({\bf 0},0)\rangle
\label{def-struc-fac}
\end{eqnarray}
of a liquid of flux lines described by the trajectories ${\bf R}_i(z)=({\bf r}_i(z),z)$ as they traverse the 
superconducting sample (here $\hat\rho({\bf r},z)=\sum_{i=1}^N\delta({\bf r}-{\bf r}_i(z))$ is the density operator 
in the ${\bf r}=(x,y)$ plane at heigh $z$, $N$ being the total number of vortices in the sample), is such that the 
partial Fourier transform 
$S({\bf q},z)=\langle\hat\rho({\bf q},z)\hat\rho(-{\bf q},0)\rangle$
is given by
\begin{eqnarray}
S({\bf q},z) = S({\bf q},z=0)\,\mbox{e}^{-\varepsilon(q)|z|/{ }T}
\label{S-Nelson-Seung}
\end{eqnarray}
where $\varepsilon(q)$ is the excitation spectrum of the equivalent boson superfluid (whose precise form 
will be given below), and $T$ is temperature (throghout this paper, we use units such that Boltzmann's constant 
$k_B=1$).
Although the Boson mapping (and other similar approaches\cite{Radzihovsky-Frey} which use the density as the basic
dynamical variable) does not make any explicit predictions for the correlations of conformational variables of flux 
lines, the exponential decay (\ref{S-Nelson-Seung}) of density correlations 
along the direction of the lines is generally {\em interpreted} as an indication that the flux line liquid is in a 
heavily entangled or ``superfluid'' state characterized by correlations of the form
\begin{eqnarray}
\langle|{\bf r}(z) - {\bf r}(0)|^2\rangle = 2D|z|\quad ,
\label{uzu0Boson}
\end{eqnarray}
with a ``diffusion'' constant $D$ of order (here $\kappa$ is the tilt modulus of the flux lines)~:
\begin{eqnarray}
D \approx \frac{{ }T}{\kappa}\quad ,
\end{eqnarray}
and by a mean projected area of the flux lines~:
\begin{eqnarray}
\langle u^2\rangle \approx DL \quad,\quad {\bf u}(z) = {\bf r}(z) - \langle {\bf r}(z)\rangle
\label{usqrdBoson}
\end{eqnarray}
which {\em diverges} with the superconducting sample thickness $L$.

Since its inception by Nelson and Seung\cite{Nelson-Seung}, the boson
mapping has had a widespread acceptance and a rather strong impact on our understanding of flux 
line liquids. It has been applied, in particular, to disorder-free vortex 
liquids\cite{Benetatos-Marchetti}, as well as to vortex liquids in presence of point\cite{Nelson-LeDoussal} 
and correlated\cite{Nelson-Vinokur} disorder. In all these studies, the boson mapping gave 
seemingly reasonable results, except at one occurrence where it was noticed by
T\"auber and Nelson\cite{Tauber-Nelson} that the boson mapping gives nonsensical results in the presence of 
splayed columnar disorder\cite{Lehrer-Nelson}.

In a recent manuscript\cite{Ettouhami}, henceforth referred to as (I), I have proposed a new approach to
study three-dimensional flux line liquids in type II superconductors. This approach, which makes contact with the standard
theory of classical fluids, is based on the separation of the flux lines'
conformal variables into center of mass (c.m.) and internal modes, and on the observation that the
interactions between flux lines may lead, under certain conditions, to the confinement of
the internal modes whose fluctuations are now bounded and no-longer
diverge with the sample thickness $L$ as in Eq. (\ref{usqrdBoson}). 
While, as mentioned above, the Boson mapping does not deal with the 
conformation variables of flux lines themselves, the theory developed in (I) predicted that the 
effective Hamiltonian of the {\em internal modes} of interacting flux lines in a vortex liquid has the form
($i$ here labels flux lines, and $N$ is the total number of vortices in the sample)
\begin{eqnarray}
H_{eff}[{\bf u}_i] = \sum_{i=1}^N\int_0^L dz\;\Big[
\frac{1}{2}\,\kappa\Big(\frac{d{\bf u}_i}{dz}\Big)^2 + \frac{1}{2}\,\mu_0 u_i^2(z)
\Big]
\end{eqnarray}
where the mass $\mu_0$ is given by
\begin{eqnarray}
\mu_0 = \frac{\rho}{2}\int d^2{\bf r}\;\nabla^2V({\bf r})g_0({\bf r}) \; .
\label{mu0}
\end{eqnarray}
In the above equation, $V$ is the interaction potential between vortices (see below, Eq. (\ref{Hamiltonian})) 
and $g_0$ is the pair distribution 
function of the two-dimensional liquid formed by the centers of mass of the flux lines. 
Because in (I) we only used a simple Taylor expansion of the total Hamiltonian 
of flux lines in terms of the ${\bf u}_i$'s, the ``mass'' term $\mu_0$ does not fully capture the effect of thermal 
fluctuations, and varies with temperature only through the (weak) variation of $g_0$. 
In particular, Eq. (\ref{mu0}) shows that the value of the ``mass'' term $\mu_0$ 
remains almost unchanged as temperature $T\to \infty$, which of course is unrealistic, since we expect the vortex 
liquid to behave as an ideal gas of freely fluctuating flux lines in this limit.

In the present paper, we generalize the method proposed in (I), and construct a variational theory of interacting flux 
lines in the vortex liquid state. This variational method, which in fact turns out to be nothing but a 
self-consistent Hartree approximation, captures the effect of thermal fluctuations in a better way and leads to a 
``renormalized'' mass term $\mu(T)$ which vanishes at high enough temperatures, as suggested by physical intuition.

Our method of approach will be as follows. In section \ref{Variational} we construct our variational approach and 
derive the expression of the renormalized $\mu(T)$. 
In section \ref{Structure-Factor}, we compare the method developed in (I) and in the present paper with the approaches 
of refs.\cite{Nelson,Nelson-Seung,Nelson-LeDoussal,Tauber-Nelson,Benetatos-Marchetti,Radzihovsky-Frey}, and discuss in 
particular the interacting structure factor $S({\bf r},z)$ derived using both methods.
Section \ref{Conclusion} contains our conclusions.

\section{Variational theory of flux line liquids}
\label{Variational}

We thus consider a liquid of interacting flux lines in a $d=(d_\perp+1)$-dimensional
superconducting sample of thickness $L$.
We here will use the following Hamiltonian\cite{Nelson-Seung,Nelson-Vinokur}
\begin{equation}
H = \sum_{n=1}^N\int_0^L dz \Big\{
\frac{1}{2}\,\kappa\Big(\frac{d{\bf r}_n}{dz}\Big)^2 \!+\! \frac{1}{2}\!\!\sum_{m(\neq n)}
V\big({\bf r}_n(z) \!-\! {\bf r}_m(z)\big) \Big\}
\label{Hamiltonian}
\end{equation}
where $\kappa$ is the tilt modulus per unit length of the flux lines,
and $V({\bf r})$ is the interaction potential between flux line elements at equal height $z$. For a uniaxial HTSC,
with both the average vortex direction and the unit vector $\hat{\bf z}$ aligned with the $\hat{\bf c}$ axis,
$\kappa\simeq\varepsilon^2\varepsilon_0$ and $V(r)=2\varepsilon_0K_0(r)/\lambda_{ab}$, where
the energy scale $\varepsilon_0$ is given by $\varepsilon_0=(\phi_0/4\pi\lambda_{ab}^2)^2$,
$\varepsilon=\lambda_{ab}/\lambda_c$ is the ratio of the 
London penetration depths in the $(ab)$ planes and in the direction of the $\hat{\bf c}$ axis, respectively;
$\phi_0=hc/2e$ is the flux quantum, and $K_0$ is a modified Bessel function\cite{Abramowitz}.
In the developments that will follow, we will find it useful to work in Fourier space, and to write down the following
decomposition of the flux line trajectories ${\bf r}_i(z,t)$ into Rouse\cite{Doi-Edwards} modes
\begin{eqnarray}
{\bf r}_{i}(z) = \sum_{n=-\infty}^\infty {\bf r}_i(q_n)\mbox{e}^{iq_n z}
\end{eqnarray}
where $q_n=2\pi n/L$, and where the Fourier coefficients ${\bf r}_i(q_n)$ are given by
\begin{eqnarray}
{\bf r}_i(q_n) = \frac{1}{L}\int_{0}^{L} dz\; {\bf r}_{i}(z)\,\mbox{e}^{-iq_n z}
\end{eqnarray}
It will also be convenient to write ${\bf r}_i(z)$ as the sum
\begin{eqnarray}
{\bf r}_i(z) = {\bf r}_{0i}(t) + {\bf u}_i(z,t)
\label{u-decomposition}
\end{eqnarray}
where ${\bf r}_{0i}(t)={\bf r}_i(q_n=0,t)$ is the c.m. position and ${\bf u}_i(z)$ the flux line
displacement with respect to the center of mass position at height $z$.

The statistical mechanics of the flux line system, Eq. (\ref{Hamiltonian}), is described by the partition function
\begin{eqnarray}
{\cal Z} & = & 
\int \prod_{i=1}^N [d{\bf r}_{i}(z)]\,\mbox{e}^{-\beta H}
\\
& = & \int\prod_{i=1}^N d{\bf r}_{0i}\int \prod_{i=1}^N [d{\bf u}_{i}(z)]\,\mbox{e}^{-\beta H}
\label{Z}
\end{eqnarray}
where $\beta=1/T$ is the inverse temperature, 
and where in going from the first to the second line we separated the path integral over every flux 
line trajectory into a path integral over all internal modes and an ordinary integral over the c.m. coordinates.
The integration measure $[d{\bf u}_{i}(z)]$ is given by
\begin{eqnarray}
[d{\bf u}_{i}(z)] = \prod_{n=1}^\infty d{\bf u}_{re}(q_n)\,d{\bf u}_{im}(q_n)
\end{eqnarray}
with ${\bf u}_{re}(q_n)$ and ${\bf u}_{im}(q_n)$ the real and imaginary parts of ${\bf u}(q_n)$, respectively.
Now, if we were able to perform the functional integrations over the 
internal modes, such an integration would give us the expression
\begin{eqnarray}
{\cal Z} = \int\prod_{i=1}^N d{\bf r}_{0i}\;\mbox{e}^{-\beta H_{eff}[\{{\bf r}_{0i}\}]}
\end{eqnarray}
which looks like the partition function of a system of $N$ ordinary classical particles with spatial coordinates
$\{{\bf r}_{0i}\}$, interacting through the Hamiltonian $H_{eff}[\{{\bf r}_{0i}\}]$, 
and whose thermodynamics can be studied by standard methods 
of statistical mechanics. Unfortunately, since the internal modes $\{{\bf u}_{i}(z)\}$ appear explicitely in the 
argument of the interaction potential $V$ in Eq. (\ref{Hamiltonian}), it is not possible, in general, to perform the 
functional integrations over these variables in closed form. In (I), we used a simple Taylor expansion of the 
interaction potential $V({\bf r}_i(z)-{\bf r}_j(z))=V({\bf r}_{0i}-{\bf r}_{0j}+{\bf u}_i(z)-{\bf u}_j(z))$
in terms of the ${\bf u}_i$'s to be able to perform the integrations in Eq. (\ref{Z}) and make predictions for 
physical observables of our flux liquid. Here, and in order to better capture the effect of thermal fluctuations, we 
want to improve on the above Taylor expansion and use a variational approach to derive the statistical mechanics of 
our model system.

In general, variational approaches to classical statistics are based on the Jensen-Peirels inequality \cite{Feynman}
\begin{eqnarray}
{\cal Z} \ge Z_v = Z_1\,\mbox{e}^{-\beta\langle H-H_1\rangle_1}
\end{eqnarray}
where $H_1$ is {\em any} Hamiltonian, 
and where the average $\langle\cdots\rangle_1$ is performed with the statistical 
weight $\exp(-\beta H_1)/Z_1$, with $Z_1=Tr(\mbox{e}^{-\beta H_1})$. 
While an arbitrary choice of $H_1$ will not necessarily lead to good results, 
a judicious choice of the 
variational Hamiltonian $H_1$ (that is simple enough so that thermal averages can be calculated and at the same time 
general enough to capture the physics of the full Hamiltonian) can lead to a variational free energy
\begin{eqnarray}
F_v & = & -T\ln Z_v
\nonumber\\
& = & -T\ln Z_1 +\langle H-H_1\rangle_1
\label{F_v}
\end{eqnarray}
which is a very good approximation of the true free energy of the sytem, and to an accurate description of the overall 
behaviour of the original model.

We now have to choose a variational Hamiltonian to approximate the Hamiltonian $H$ of equation (\ref{Hamiltonian}). 
In what follows, we shall assume that we can write for our variational Hamiltonian $H_1$ a decomposition of the form~:
\begin{eqnarray}
H_1[\{{\bf r}_{0i}\},\{{\bf u}_i(z)\}]  = {H}_0[\{{\bf r}_{0i}\}]  +  H_u[\{{\bf u}_i(z)\}]
\label{defH1}
\end{eqnarray}
In the above decomposition,
${H}_0[\{{\bf r}_{0i}\}]={H}_0({\bf r}_{01},\ldots,{\bf r}_{0N})$ is an effective
variational Hamiltonian describing the interactions of the c.m. modes, while $H_u[\{{\bf u}_i(z)\}]$ 
describes the elasticity and mutual interactions of the internal modes of flux lines. For $H_u[\{{\bf u}_i(z)\}]$, we 
shall use the following Gaussian approximation (here and in the following, we sum over repeated
indices unless otherwise indicated)~:
\begin{equation}
H_u = \frac{1}{2L^2}\sum_{i,j}\int \!\! dz\int \!\! dz'\;[G^{-1}(z-z')]_{ij}
{\bf u}_i(z)\cdot{\bf u}_j(z')
\label{Hureal}
\end{equation}
Throughout the rest of the paper, it will be understood that integrals over the $z$ and $z'$ variables run from $0$ 
to $L$. The above expression can be rewritten in Fourier space in the form
\begin{equation}
H_u  =  \frac{1}{2}\,\sum_{i,j}\sum_{n\neq 0}\;[G^{-1}(q_n)]_{ij} 
{\bf u}_{i}(q_n)\cdot{\bf u}_{j}(-q_n) 
\label{HuFourier} 
\end{equation}   
In Eqs. (\ref{Hureal})-(\ref{HuFourier}), $[G^{-1}(q_n)]_{ij}$ is an $N\times N$ 
matrix of variational parameters describing the interactions of the internal modes of flux lines
\cite{Remark}.
In the boson language, the decomposition (\ref{defH1}) of the total Hamiltonian into c.m. and internal modes pieces can 
be seen as the the generalization of the Feynman-Kleinert variational 
method \cite{Feynman-Kleinert} for single quantum particles in imaginary time
to an assembly of interacting quantum particles. 
Both quantities ${H}_0$ and $[G^{-1}(q_n)]_{ij}$ will be determined variationally, and one can in fact show, following 
ref.  \onlinecite{Mezard}, that the Gaussian approximation (\ref{HuFourier}), with the optimal choice for the 
propagator $[G^{-1}(q_n)]_{ij}$ to be detemined below, becomes {\em exact} in the limit of a large 
number of components of the displacement field $\{{\bf u}_{i}(z)\}$.

Using the Hamiltonian $H_1$ above, Eqs.(\ref{defH1})-(\ref{HuFourier}), to evaluate the variational 
free energy in Eq. (\ref{F_v}), we find
\end{multicols}
\begin{eqnarray}
F_v & = & -T\ln Z_1 + \frac{1}{Z_1}\int\prod_{i=1}^N d{\bf r}_{0i}\;\Phi({\cal G})\,\mbox{e}^{-\beta{H}_0}
\Big[
\frac{1}{2}\sum_{i,j}\sum_{n\neq 0} d_\perp TL
\Big([G_0^{-1}(q_n)]_{ij} - [G^{-1}(q_n)]_{ij}\Big)\,G_{ij}(q_n) +
\nonumber\\
& + & \frac{1}{2}\,L\;\sum_{i,j}\int_{\bf q} V({\bf q})\,\mbox{e}^{i{\bf q}\cdot
({\bf r}_{0i}-{\bf r}_{0j})}\,\mbox{e}^{-\frac{1}{2}q^2C_{ij}} - {H}_0[\{{\bf r}_{0i}\}]
\Big]
\label{Fvcomplete}
\end{eqnarray}
where
\begin{eqnarray}
Z_1 & = & \int\prod_{i=1}^N d{\bf r}_{0i}\int [d{\bf u}_i(z)]\;\mbox{e}^{-\beta {H}_0}\exp\Big(
-\frac{\beta}{2}\,\sum_{i,j}\sum_{n\neq 0} (G^{-1}(q_n))_{ij}{\bf u}_i(q_n)\cdot{\bf u}_j(-q_n)
\Big)
\nonumber\\
& = & \int\prod_{i=1}^N d{\bf r}_{0i}\;\Phi({\cal G})\,\mbox{e}^{-\beta {H}_0}
\label{Z1fin}
\end{eqnarray}
\begin{multicols}{2}
and
\begin{eqnarray}
\Phi({\cal G}) = \prod_{n=1}^\infty\Big(
\frac{2\pi}{\mbox{det}(T{\cal G}(q_n))}
\Big)^{d_\perp} \quad,
\end{eqnarray}
and where ${\cal G}(q_n)$ denotes the $N\times N$ matrix $G_{ij}(q_n)$.
The correlation function $C_{ij}$ is on the other hand given by
\begin{eqnarray}
C_{ij}=\frac{1}{d_\perp}\;\langle [{\bf u}_i(z) - {\bf u}_j(z)]^2\rangle
\end{eqnarray}
Variation of the above expression with respect to the c.m. Hamiltonian ${H}_0$ shows that $F_v$ is minimal for 
the particular choice
\end{multicols}
\begin{eqnarray}
\tilde{H}_0 = \frac{1}{2}\sum_{i,j}\sum_{n\neq 0} d_\perp TL
\Big([G_0^{-1}(q_n)]_{ij} - [G^{-1}(q_n)]_{ij}\Big)\,G_{ij}(q_n)
+\frac{1}{2}\,L\;\sum_{i,j}\int_{\bf q} V({\bf 
q})\,\mbox{e}^{i{\bf q}\cdot
({\bf r}_{0i}-{\bf r}_{0j})}\,\mbox{e}^{-\frac{1}{2}q^2C_{ij}}
\label{H10}
\end{eqnarray}
\begin{multicols}{2}
The trial free energy $F_v$ of Eq. (\ref{Fvcomplete}) thus reduces to
\begin{eqnarray}
F_v = -T\ln Z_1
\label{Fvfinal}
\end{eqnarray}
where now in the expression of $Z_1$, Eq. (\ref{Z1fin}), the expression 
(\ref{H10}) of $\tilde{H}_0$ should be used for ${H}_0$.

Further minimization of $F_v$ with respect to $G_{ij}(q_n)$ leads to the following results (Appendix A)
\end{multicols}
\begin{mathletters}
\begin{eqnarray}
{[\tilde{G}^{-1}(q_n)]}_{ij} & = & \kappa q_n^2 - \frac{\rho}{d_\perp}\int_{\bf q} q^2V({\bf q})g_0({\bf q})
\mbox{e}^{-q^2 \langle u^2\rangle/d_\perp} \quad \mbox{for} \quad i = j
\label{resGa}
\\
& = & \frac{\rho}{d_\perp(N-1)}\int_{\bf q} q^2V({\bf q})g_0({\bf q})
\mbox{e}^{-q^2 \langle u^2\rangle/d_\perp} \quad \mbox{for} \quad i\neq j
\label{resGb}
\end{eqnarray}
\end{mathletters}
\begin{multicols}{2}
\noindent
where the tilde indicates that the inverse propagator $\tilde{G}^{-1}(q_n)$ has been averaged over all 
possible 
configurations of the c.m. positions $\{{\bf r}_{0i}\}$ compatible with a liquid structure.


The integral on the right hand side of Eq. (\ref{resGb}) being finite and independent of the total number of vortices 
$N$, we see that the off-diagonal elements ${[\tilde{G}^{-1}(q_n)]}_{i\neq j}$ vanish in the thermodynamic 
$N\to\infty$ limit, in which case the internal modes elastic propagator is diagonal and given by:
\begin{eqnarray}
[\tilde{G}^{-1}(q_n)]_{ij} = L(\kappa q_n^2 + \mu)\;\delta_{ij}
\end{eqnarray}
with $\delta_{ij}$ the Kronecker delta symbol and where $\mu$ is given by:
\begin{eqnarray}
\mu = -\frac{\rho}{d_\perp}\int_{\bf q} q^2V({\bf q})g_0({\bf q})  
\mbox{e}^{-q^2 \langle u^2\rangle/d_\perp}
\label{defmu}
\end{eqnarray}
In the above equation, the mean square width of a flux line
\begin{eqnarray}
\langle u^2\rangle & = & d_\perp T\sum_{n\neq 0}\tilde{G}(q_n)
\nonumber\\
& = & \frac{d_\perp T}{4\sqrt{\kappa\mu}}
\label{usqrd}
\end{eqnarray}
depends on $\mu$. Thus we see that Eq. (\ref{defmu}) is in fact a self-consistent equation for the ``mass'' 
coefficient $\mu$. From this last equation, we also see that $\mu$ depends on
the density $\rho$ of flux lines, not only through the prefactor in front of the integral, but also through the pair 
distribution function $g_0({\bf r})$. Through this last quantity, $\mu$ also depends on temperature and on higher, 
nontrivial correlations between the positions of flux lines in the vortex liquid. The strongest dependence of $\mu$
on temperature, however, occurs through the ``Debye-Waller'' factor $\mbox{e}^{-q^2 \langle u^2\rangle/d_\perp}$
which was missing in our more elementary Taylor expansion in (I), and which describes the effect of internal 
fluctuations of flux lines.

\begin{center}   
\begin{figure}[b]
\includegraphics[scale=0.40]{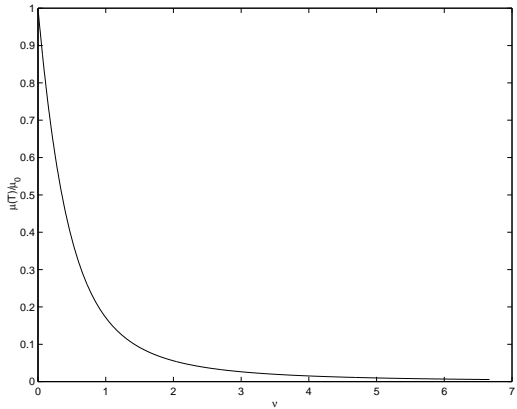}
\caption[]
{Plot of $\mu(T)/\mu_0$ versus $\nu={\alpha T}/{4a^2\sqrt{\kappa\mu_0}}$.
}\label{plotmuvsT}
\end{figure}
\end{center}

From Eq. (\ref{defmu}), it is not possible to derive a general expression of the coefficient $\mu$
without making assumptions about the analytic form of the pair correlation function $g_0({\bf r})$. 
Here, we shall place ourselves in the case of a dilute flux liquid and use the
analytical ansatz we proposed in (I) for the pair correlation function $g_0({\bf r})$, namely~: 
\begin{eqnarray}
g_0({\bf r}) = 1 - \eta\exp(-{\alpha}r^2/a^2)
\label{g0}
\end{eqnarray}
where $\alpha$ is a constant of order unity, and $0<\eta<1$. The numerical constant $\eta$ quantifies the correlations 
between c.m. positions of flux lines. It is close to unity when flux lines are strongly anti-correlated due to the 
repulsive interactions between their surrounding supercurrents, and close to zero in situations where there is 
considerable cutting and crossing of flux lines. Using the above ansatz for $g_0(r)$, we obtain the following 
expression for the ``mass'' $\mu$ of the internal modes as a function of $T$:
\begin{equation}
\mu(T) = \mu_0\;\left( \sqrt{1+\Big(\frac{\alpha T}{4a^2\sqrt{\kappa\mu_0}}\Big)^2} - 
\Big(\frac{\alpha T}{4a^2\sqrt{\kappa\mu_0}}\Big)\right)^2 
\label{muvsT}
\end{equation}
where $\mu_0=\mu(T=0)=2\eta\pi\rho\varepsilon_0/d_\perp$. Fig.\ref{plotmuvsT} shows a plot of $\mu(T)/\mu_0$ as a 
function of the variable $\nu={\alpha T}/{4a^2\sqrt{\kappa\mu_0}}$.

From Eq. (\ref{muvsT}), it is not difficult to see that the above expression has the 
limiting value $\mu(T)\to\mu_0$ when $T\ll a^2\sqrt{\kappa\mu_0}$, while at higher temperatures 
$T\gg a^2\sqrt{\kappa\mu_0}$, $\mu(T)\simeq \mu_0(4a^2\sqrt{\kappa\mu_0}/\alpha T)$ and goes to zero like $1/T$.
We thus see that $\langle u^2\rangle$ in Eq. (\ref{usqrd}) will vary with temperature like $T^{3/2}$. More generally, 
any nontrivial form of the pair distribution function $g_0(r)$ in Eqs. (\ref{resGa})-(\ref{resGb}) will result in a 
nontrivial temperature dependence of $\mu(T)$ and hence of the mean projected area of a flux line $\langle u^2\rangle$.
This temperature variation of $\langle u^2\rangle$ can be measured in numerical simulations and is actually a 
good way to test the predictions of the present paper. Note also that 
\noindent as $T$ becomes very large, $\mu(T)$
vanishes and we recover a liquid of freely fluctuating flux lines with $\langle u^2\rangle$ diverging with the sample 
thickness $L$.


\section{The interacting structure factor of flux line liquids}
\label{Structure-Factor}

We now turn our attention to a comparison and a discussion of the results obtained by the approach developed in (I) and 
the present paper, and the Boson-mapping type of approach. Since this last type of approach only makes predictions 
for correlations of coarse-grained variables, such as the density, and makes no predictions for microscopic 
conformation variables of the flux lines, we here shall focus on comparing the interacting 
structure factor $S({\bf r},z)$ obtained through both methods.

Let us first start by deriving the structure factor of a liquid of flux lines from our variational approach.
This can be done in a straightforward way, with the result (Appendix B)~:
\begin{eqnarray}
S({\bf q},z) = \rho^2 g_0({\bf q}) \,\mbox{e}^{-q^2\langle u^2\rangle/d_\perp} +
\rho\mbox{e}^{-\frac{1}{2d_\perp}q^2\phi(z)}
\label{Sgen}
\end{eqnarray}
Let us insist that this result is actually very general, and only requires that an approximate decomposition of the
form (\ref{defH1}) can be written for the Hamiltonian of the flux lines in terms of the conformational degrees of 
freedom of flux lines, and that the internal modes of flux lines 
obey a Gaussian distribution such as the one implied by Eq. (\ref{HuFourier}). Whith these assumptions, the mean square 
relative displacement in Eq. (\ref{Sgen}), for an arbitrary form of the propagator $G_{ij}(q_n)$ in 
Eq.~(\ref{HuFourier}), is given by (here $G(q_n)$ is the diagonal element $G_{ii}(q_n)$)~:
\begin{eqnarray}
\phi(z) & = & \langle[{\bf u}_i(z)-{\bf u}_i(0)]^2\rangle
\nonumber\\
& = & 2d_\perp { }T \sum_{n\neq 0} G(q_n)\,\big[1 - \cos(q_nz)\big]
\label{def-phi}
\end{eqnarray}
while the mean projected area of a given flux line is given by
\begin{eqnarray}
\langle u^2\rangle = d_\perp { }T \sum_{n\neq 0} G(q_n)
\label{usqrdgen}
\end{eqnarray}

Having derived the structure factor through our approach, we want to compare it to the corresponding quantity derived
through the Boson mapping and similar approaches.
It has been shown in reference\cite{Nelson-Seung} using the boson mapping that this last quantity, in Fourier space, is
given by~:
\begin{eqnarray}
S({\bf q},q_z) = \frac{\rho { }Tq^2/{\kappa}}{q_z^2+\varepsilon^2({\bf q})/T^2}
\label{strucfac}
\end{eqnarray}
where the excitation spectrum of the equivalent bosons has the usual form
\begin{eqnarray}
\frac{\varepsilon({\bf q})}{{ }T} = \Big[
\Big(\frac{{ }Tq^2}{2\kappa}\Big)^2 + \frac{\rho q^2V({\bf q})}{\kappa}
\Big]^{1/2}
\label{spectrum}
\end{eqnarray}
From equation (\ref{strucfac}), it is not difficult to obtain that the partial Fourier
transform $S({\bf q},z)$ has the the following expression
\begin{eqnarray}
S({\bf q},z) & = &\int_{-\infty}^\infty \frac{dq_z}{2\pi}\;S({\bf q},q_z)\,\mbox{e}^{iq_zz}
\nonumber\\
& = &
\frac{\rho\,\mbox{e}^{-\varepsilon({\bf q})|z|/{ }T}}
{\sqrt{1+4{\rho}V(q){\kappa}/({ }Tq)^2}}
\label{Sboson}
\end{eqnarray}

We first note that the above analytic form of the structure factor, Eq. (\ref{Sgen}), and the hydrodynamic result
(\ref{Sboson}) cannot be brought to agreement. In other words, {\em no} Gaussian approximation of the
form (\ref{HuFourier}) can produce a structure factor of the form (\ref{Sboson}).

As we already mentioned in the introduction, it has already been noticed \cite{Tauber-Nelson,Lehrer-Nelson}
that the boson mapping gives nonsensical results when applied to flux liquids in the presence of splayed columnar disorder. 
In Appendix B, we comment on the definition of the chemical potential of flux lines 
in the boson mapping, which appears to us a little problematic. While this may be considered to be a rather 
``cosmetic'' issue, in what follows we show that the 
boson mapping yields other unphysical results (in addition to those aleady noticed in ref.\cite{Tauber-Nelson})
even in the simplest case of pure, disorder free vortex liquids.
To this end, let us make the following two observations regarding the analytic form of $S({\bf q},z)$ in the boson 
scheme~:

(i) By taking the limit $\kappa\to \infty$ in equation (\ref{strucfac}), one should be able to
obtain the structure factor of a liquid of rigid flux lines, which we expect to be independent of $z$.
Taking the above limit in equation (\ref{Sboson}), we see that the $z$-dependence does indeed cancel out, 
since $\varepsilon({\bf q})\to 0$ as $\kappa\to\infty$. However, the prefactor 
\begin{eqnarray}
S({\bf q},z=0) = \frac{\rho { }Tq^2/\kappa}{2\big(\varepsilon({\bf q})/{ }T\big)}  
= \frac{\rho}{\sqrt{1+4{\rho}V(q){\kappa}/({ }Tq)^2}}
\nonumber
\end{eqnarray}
also goes to zero in the 
limit $\kappa\to \infty$,
which means that the whole structure factor $S({\bf q},z=0)$ 
will vanish as $\kappa$ gets very large, which is not exactly what we 
would expect for a liquid of rigid flux lines. In contrast to this behaviour, the structure factor derived within our 
variational perturbation theory is well-behaved in the limit $\kappa\to\infty$, and it is not difficult to see from 
Eqs. (\ref{Sgen})-(\ref{usqrdgen}) that we obtain, for an arbitrary choice of $G_{ij}(q_n)$ (with either $\mu=0$ or a 
nonzero $\mu$), in the above limit~:
\begin{equation}
S({\bf q},z) = \rho\,\big[ 1 + \rho\,g_0(r)\big]
\end{equation}
which is the correct result for a liquid of rigid hard rods.

(ii) For an ``ideal gas'' of non-interacting flux lines, the mean projected area $\langle u^2\rangle$ of a given flux 
line is given by 
\begin{eqnarray}
\langle u^2\rangle = \frac{d_\perp { }TL}{12\kappa}
\label{usqrdfree}
\end{eqnarray}
and the mean square displacement $\phi(z)$ is given by
\begin{eqnarray}
\phi(z) = \frac{2{ }T}{\kappa}\,|z|
\label{phi-free}
\end{eqnarray}
Using these expressions in Eq. (\ref{Sgen}) above, one finds, for non-interacting flux lines
in the thermodynamic ($L\to\infty$) limit, 
\begin{eqnarray}
S_{id.}({\bf q},z) =  \rho \,\exp\Big(-\frac{{ }T q^2}{2{\kappa}}\,|z|\Big)
\label{S0}
\end{eqnarray}
Comparing this expression to the expression (\ref{Sboson})
of the structure factor of an interacting flux liquid in the boson mapping
and given the expression (\ref{spectrum}) of the Boguliubov spectrum, we see that,
at any given value of ${\bf q}$, $S({\bf q},z)$ decays more rapidly as a function of $z$ than its free counterpart 
$S_{id.}({\bf q},z)$. This result is very surprizing, to say the least, as it is very difficult to imagine a flux line 
assembly where the density decorrelates on shorter length scales 
and vortices fluctuate more strongly than in an ``ideal gas'' of freely fluctuating flux 
lines. Indeed, the effect of the repulsive interactions between flux lines is to reduce the fluctuations of vortices, 
as is evidenced at the freezing transition: in the absence of interactions, a liquid of vortices would never freeze 
into a vortex solid and would continue to have large fluctuations, Eqs. (\ref{usqrdfree})-(\ref{phi-free}), down to 
$T=0$. Going back to equation (\ref{Sgen}), it is not difficult to verify that our result for the 
structure factor does {\em not} exhibit the above behaviour. With the particular choice (\ref{g0}) for the pair 
distribution function $g_0(r)$, we obtain \cite{Ettouhami}
\begin{eqnarray}
S({\bf q},z) & = & \rho^2 g_0({\bf q})\mbox{e}^{-\frac{q^2T}{4\sqrt{\kappa\mu(T)}}} +
\nonumber\\
& + & \rho\,\exp\Big(
-\frac{Tq^2}{2\sqrt{\kappa\mu(T)} }\big( 1\!- \!\mbox{e}^{- \sqrt{\mu(T)/\kappa}|z|}\big)
\Big) \label{S(z)}
\end{eqnarray}
which has the following, nonvanishing limit as $z\to\infty$
\begin{equation}
S({\bf q},z\to\infty) \!=\!  \rho^2 g_0({\bf q})\mbox{e}^{-\frac{q^2T}{4\sqrt{\kappa\mu(T)}}} +
\rho\exp\Big(
-\frac{Tq^2}{2\sqrt{\kappa\mu(T)} }
\Big) \label{S(zinf)}
\end{equation}
indicating that the flux line densities $\hat\rho({\bf q},z)$ at both sides $z=0$ and $z=L$ of a superconducting 
sample are not very much decorrelated, in agreement with the experimental findings of ref. \onlinecite{Yoon}.
In this reference, the authors have performed a two-sided
decoration experiment on both sides of a superconducting sample, and 
observed the same pattern of vortices in each case. If the structure factor obeyed the boson mapping result, Eq. 
(\ref{Sboson}), the densities $\hat\rho({\bf q},0)$ and $\hat\rho({\bf q},L)$ would have been completely 
decorrelated and very different from one another. The fact that both density patterns in the experiments of  
ref. \onlinecite{Yoon} look exactly the same might be
evidence of the fact that there is a strong 
correlation between $\hat\rho({\bf q},0)$ and $\hat\rho({\bf q},L)$, and that 
$S({\bf q},z)$ is finite in the limit $z\to\infty$.

In fact, one can even attempt a slightly more quantitative comparison of our results, Eqs. 
(\ref{S(z)})-(\ref{S(zinf)}), and the 
experimental findings of ref. \onlinecite{Yoon}. From Eqs. (\ref{S(z)})-(\ref{S(zinf)}), it is not difficult to see 
that the ratio $S({\bf q},z=0)/S({\bf q},z=L)$ is of order unity for ${\bf q}$ not too close to 0, and 
is such that
\begin{eqnarray}
\frac{S({\bf q},z=0)}{S({\bf q},z=L)} \to 1 \quad \mbox{as} \quad {\bf q}\to 0
\end{eqnarray}
in agreement with the decoration experiments of Yoon et {\em al}. It would be interesting to do a more detailed 
analysis of the experimental results in light of the present model (taking into account nonlocal 
elasticity\cite{Brandt}), as such 
an analysis may lead to values of the elastic constants of the flux lattice (in particular, of the tilt modulus 
$\kappa$) which are closer to the conventional values of these quantities (recall that the analysis of the 
experiments of ref. \onlinecite{Yoon} using the boson mapping yields elastic constants which are three to four orders 
of magnitude below what is expected from the standard theory of elacticity of vortices in uniaxial type II 
superconductors\cite{Brandt,Sardella,Blatter}).

\section{Conclusion}
\label{Conclusion}

In summary, in this paper we have extended the approach developed in (I) to better take into account the effect of 
thermal fluctuations of flux lines in vortex liquids. Using a variational Hartree approximation,  
we have shown that thermal fluctuations strongly reduce the ``mass'' term which confines the fluctuations of the 
internal modes of flux lines. We have also argued that our approach,
which is based on the use of the conformation variables $\{{\bf r}_n(z)\}$ as the 
{\em true} dynamical variables in terms of which all (Gaussian) averages are taken, 
yields physically more reasonable results than the boson 
mapping \cite{Nelson-Seung,Nelson-LeDoussal,Tauber-Nelson} or other hydrodynamic approaches\cite{Radzihovsky-Frey} 
which, in contrast, use the density $\hat\rho({\bf r},z)$ as the basic dynamical variable of the system.
In (I), we had argued that a careful numerical measurement of the pair distribution function for the c.m. mode 
$g_0(r)$ could shed more light on the confining mass term $\mu_0$ in a vortex liquid. Here, we have shown that a 
numerical measurement of the temperature dependence of the mean square projected area of flux lines can also be used to 
test the predictions of the present approach. In particular, a nonlinear temperature dependence of $\langle u^2\rangle$
could be the signature of a weakly entangled state, where vortex fluctuations, as measured by
$\langle u^2\rangle$, do {\em not} diverge with the sample thickness $L$.

It is worth noting that the considerations of (I) and of the present work can be easily extended 
to the dynamics of flux liquids. This will be the subject of a future contribution \cite{unpublished}.

\acknowledgments

The author acknowledges support by the NSF through grant DMR--9625111 and by the Lucile and David Packard Foundation.

\section{Appendix A~: Minimization of the variational free energy}

In this Appendix, we present a few details of the minimization of the free energy of Eq. (\ref{Fvfinal}) with respect 
to the propagator of internal fluctuations $G_{kl}(q_m)$. Taking the derivative of Eq. (\ref{Fvfinal})
with respect to this last quantity amounts to computing the following expression
\begin{eqnarray}
\frac{\partial F_v}{\partial G_{kl}(q_m)}= \frac{1}{Z_1}\int\prod_{i=1}^N \! d{\bf r}_{0i}\;
\frac{\partial H_{eff}}{\partial G_{ij}(q_n)} \;\mbox{e}^{-\beta H_{eff}[\{{\bf r}_{0i}\}]}
\label{derFv}
\end{eqnarray}
with the effective Hamiltonian of the c.m. modes:
\begin{eqnarray}
H_{eff} = \tilde{H}_0 - { }T\ln\Phi({\cal G})\quad ,
\label{Heff}
\end{eqnarray}
and the partition function
\begin{eqnarray}
Z_1 = \int\prod_{i=1}^N \!\!d{\bf r}_{0i}\;\mbox{e}^{-\beta H_{eff}[\{{\bf r}_{0i}\}]}\;.
\end{eqnarray}
The derivative of the first term in $H_{eff}$ can be found in a straightforward way, and gives~:
\end{multicols}
\begin{eqnarray}
\frac{\partial \tilde{H}_0}{\partial G_{kl}(q_m)}= LTd_\perp\kappa q_m^2\,\delta_{kl} -
\frac{1}{2}\,TL\,\sum_{i,j}\int_{\bf q} q^2V({\bf q})\mbox{e}^{i{\bf q}\cdot[{\bf r}_{0i}-{\bf r}_{0j}]}
\;\mbox{e}^{-\frac{1}{2}k^2C_{ij}}\;\big[(\delta_{ik}+\delta_{jk})\delta_{kl} - 2\delta_{ik}\delta_{jl}\big]\; .
\label{der1}
\end{eqnarray}
On the other hand, the derivative of the second term in Eq. (\ref{Heff}) can be taken in a standard 
way, with the result 
\begin{eqnarray}
\frac{\partial}{\partial G_{kl}(q_m)}\,\ln\Phi({\cal G}) = Ld_\perp\,[G^{-1}(q_m)]_{kl}
\end{eqnarray}

%
%

With hindsight from the results of (I), we can expect $\langle {\bf u}_i(z)\cdot{\bf u}_j(z)\rangle\simeq 0$, so that
$$C_{ij}=\langle[{\bf u}_i(z)-{\bf u}_j(z)]^2\rangle/d_\perp \simeq 2\langle u^2\rangle/d_\perp \;.$$
Using this last approximation into Eq. (\ref{der1}), and averaging over the c.m. positions in Eq.~(\ref{derFv}), 
using the fact that
\begin{eqnarray}
\frac{1}{Z_1}\int\prod_{i=1}^N \! d{\bf r}_{0i}\;
\mbox{e}^{i{\bf q}\cdot({\bf r}_{0i}-{\bf r}_{0j})}
\;\mbox{e}^{-\beta H_{eff}[\{{\bf r}_{0i}\}]} & = &\frac{\rho^2}{N(N-1)}\int d{\bf r}_{0i}\,d{\bf r}_{0j}\;
g_0({\bf r}_{0i}-{\bf r}_{0j})\,\mbox{e}^{i{\bf q}\cdot({\bf r}_{0i}-{\bf r}_{0j})}
\nonumber\\
& = &\frac{\rho}{N-1}\;g_0({\bf q})
\end{eqnarray}
we immediately arrive at Eq. (\ref{resGa})-(\ref{resGb}) of the text.

\begin{multicols}{2}

\section{Appendix B~: Structure factor of the flux liquid}

In this appendix, we show some details of the derivation of the structure factor
\begin{eqnarray}
S({\bf r},z;{\bf r}',z') =\langle\hat\rho({\bf r},z)\hat\rho({\bf r}',z')\rangle
\end{eqnarray}
where the statistical average $\langle \cdots\rangle$ is taken with the statistical weight
$\exp(-\beta H)/{\cal Z}$, with $H$ the Hamiltonian defined in Eq. (\ref{defH1}) and ${\cal Z}=Z_0Z_u$, with
$Z_0=\mbox{Tr}(\mbox{e}^{-\beta \tilde{H_0}})$ and $Z_u=\mbox{Tr}(\mbox{e}^{-\beta {H_u}})$.
We expect flux liquids at equilibrium to be translationally invariant, and the structure factor
to depend only on the relative coordinates $({\bf r}-{\bf r}')$ and $(z-z')$, {\em i.e.}
$S({\bf r},z;{\bf r}',z')=S({\bf r}-{\bf r}',z-z')$. As a consequence, we have for the Fourier transform of the
density-density correlation function~:
\begin{equation}
\langle\hat\rho({\bf q},q_z)\hat\rho({\bf q}',q_z')\rangle =
(2\pi)^d\delta({\bf q}+{\bf q}')\delta(q_z+q_z')\;S({\bf q},q_z)
\label{S(q)}
\end{equation}
In the following, we shall be interested in the quantity
\begin{eqnarray}
S({\bf q},q_z) = \frac{1}{LL_\perp^{d_\perp}}\,\langle\hat\rho({\bf q},q_z)\hat\rho(-{\bf q},-q_z)\rangle
\end{eqnarray}
where we used equation (\ref{S(q)}) above and the fact that
$(2\pi)^d\delta({\bf q}={\bf 0})\delta(q_z=0)\equiv LL_\perp^{d_\perp}$ in the limit
$L,L_\perp\to\infty$. We have~:
\end{multicols}
\begin{eqnarray}
\langle\hat\rho({\bf q},q_z)\hat\rho(-{\bf q},-q_z)\rangle & = &
\int d{\bf r}dz\int d{\bf r}'dz'\;\langle\hat\rho({\bf r},z)\hat\rho({\bf r}',z')\rangle\;
\mbox{e}^{-i{\bf q}\cdot({\bf r}-{\bf r}')}\mbox{e}^{-iq_z(z-z')}
\nonumber\\
& = & \sum_{i=1}^N\sum_{j=1}^N\int d{\bf r}dz\int d{\bf r}'dz'\;
\langle\delta({\bf r}-{\bf r}_i(z))\delta({\bf r}'-{\bf r}_j(z'))\rangle\;
\mbox{e}^{-i{\bf q}\cdot({\bf r}-{\bf r}')}\mbox{e}^{-iq_z(z-z')}
\nonumber\\
&=&\sum_{i=1}^N\sum_{j\neq i}\int dz\int dz'\;
\langle\mbox{e}^{-i{\bf q}\cdot({\bf r}_{0i}-{\bf r}_{0j})}\rangle_0
\langle\mbox{e}^{-i{\bf q}\cdot({\bf u}_{i}(z)-{\bf u}_{j}(z'))}\rangle_1\;\mbox{e}^{-iq_z(z-z')} +
\nonumber\\
&+& \sum_{i=1}^N\int dz\int dz'
\langle\mbox{e}^{-i{\bf q}\cdot({\bf u}_{i}(z)-{\bf u}_{j}(z'))}\rangle_1\;\mbox{e}^{-iq_z(z-z')}
\nonumber\\
&=&\sum_{i=1}^N\sum_{j\neq i}\int dz\int dz'\;
\langle\mbox{e}^{-i{\bf q}\cdot({\bf r}_{0i}-{\bf r}_{0j})}\rangle_0
\;\mbox{e}^{-\frac{1}{2}q_\alpha q_\beta
\langle[u_{i,\alpha}(z)-u_{j,\alpha}(z')][u_{i,\beta}(z)-u_{j,\beta}(z')]\rangle_1}  
\;\mbox{e}^{-iq_z(z-z')} +
\nonumber\\
&+& \sum_{i=1}^N\int dz\int dz'
\mbox{e}^{-\frac{1}{2}q_\alpha q_\beta
\langle[u_{i,\alpha}(z)-u_{i,\alpha}(z')][u_{i,\beta}(z)-u_{i,\beta}(z')]\rangle_1}
\;\mbox{e}^{-iq_z(z-z')}
\label{interm-1}
\end{eqnarray}
In the above equations, $\langle\cdots\rangle_0$ and $\langle\cdots\rangle_u$ denote averages over the c.m. and 
internal modes, with statistical weights $\exp(-\beta {H}_0)$ and $\exp(-\beta H_{u})$ respectively
(${H}_0$ and $H_{u}$ are the Hamiltonians given in equations (\ref{H10}) and (\ref{HuFourier}) of the text.
In our variational model, we find that the
internal degrees of freedom ${\bf u}_i(z)$ and ${\bf u}_j(z)$ belonging to two different lines $i\neq j$ are
decoupled. We therefore can write, for $i\neq j$~:
\begin{eqnarray}
\langle[u_{i,\alpha}(z)-u_{j,\alpha}(z')][u_{i,\beta}(z)-u_{j,\beta}(z')]\rangle_1
&=& \langle u_{i,\alpha}(z)u_{i,\beta}(z) + u_{j,\alpha}(z')u_{j,\beta}(z')\rangle_1
\nonumber\\
& = & \frac{2\delta_{\alpha\beta}}{d_\perp}\langle u^2\rangle
\end{eqnarray}
where, in going from the first to the second line, we used the fact that
$\langle u_{i,\alpha}u_{i,\beta}\rangle_1=\delta_{\alpha,\beta}\langle u_{i,\alpha}^2\rangle$.
Equation (\ref{interm-1}) becomes
\begin{eqnarray}
\langle\hat\rho({\bf q},q_z)\hat\rho(-{\bf q},-q_z)\rangle & = &
\sum_{i=1}^N\sum_{j\neq i}\int dz\int dz'\;
\langle\mbox{e}^{-i{\bf q}\cdot({\bf r}_{0i}-{\bf r}_{0j})}\rangle_0
\mbox{e}^{-\frac{1}{d_\perp}q^2\langle u^2\rangle}
\;\mbox{e}^{-iq_z(z-z')} +
\nonumber\\
&+& \sum_{i=1}^N\int dz\int dz'
\mbox{e}^{-\frac{1}{2}q_\alpha q_\beta
\langle[u_{i,\alpha}(z)-u_{i,\alpha}(z')][u_{i,\beta}(z)-u_{i,\beta}(z')]\rangle_1}
\;\mbox{e}^{-iq_z(z-z')}
\label{interm-2}
\end{eqnarray} 
Using the fact that
$\int dz\!\int dz' \;\mbox{e}^{-iq_z(z-z')} = L^2\delta_{q_z,0}$,
and noticing that
\begin{eqnarray}
\sum_{i=1}^N\sum_{j\neq i}\langle \mbox{e}^{-i{\bf q}\cdot{\bf r}_{0i}}
\mbox{e}^{-i{\bf q}'\cdot{\bf r}_{0j}}\rangle_0 =
(2\pi)^{d_\perp}\delta({\bf q}+{\bf q}')\;\rho^2\,g_0({\bf q})
\end{eqnarray}
which gives us here (with ${\bf q}'=-{\bf q}$)
\begin{eqnarray}
\sum_{i=1}^N\sum_{j\neq i}\langle \mbox{e}^{-i{\bf q}\cdot{\bf r}_{0i}}
\mbox{e}^{i{\bf q}\cdot{\bf r}_{0j}}\rangle_0 =
(2\pi)^{d_\perp}\delta({\bf q}=0)\;\rho^2\,g_0({\bf q})
= L_\perp^{d_\perp}\rho^2 g_0({\bf q})
\end{eqnarray}
we finally obtain~:
\begin{eqnarray}
\langle\hat\rho({\bf q},q_z)\hat\rho(-{\bf q},-q_z)\rangle & = &
L^2\delta_{q_z,0}\;L_\perp^{d_\perp}\,\rho^2g_0({\bf q})\;
\mbox{e}^{-\frac{1}{d_\perp}q^2\langle u^2\rangle}
+ \sum_{i=1}^N\int dz\int dz'
\mbox{e}^{-\frac{1}{2d_\perp}q^2
\langle[{\bf u}_{i}(z)-{\bf u}_{i}(z')]^2\rangle_1}
\;\mbox{e}^{-iq_z(z-z')}
\label{interm-3}
\end{eqnarray}
where we used the fact that
$\langle[u_{i,\alpha}(z)-u_{i,\alpha}(z')][u_{i,\beta}(z)-u_{i,\beta}(z')]\rangle_u =
\frac{\delta_{\alpha,\beta}}{d_\perp}\;\langle[{\bf u}_i(z)-{\bf u}_i(z')]^2\rangle_u$.
The partial Fourier transform of this last expression with respect to $q_z$ leads directly to expression 
(\ref{Sgen}) of the text.

\begin{multicols}{2}

\section{Appendix C~: Chemical potential of a flux line liquid}

In this appendix, we briefly comment on the ``chemical potential'' of a flux line liquid
adopted in references\cite{Nelson-Seung,Nelson-LeDoussal,Nelson-Vinokur,Tauber-Nelson}, which seems to us
to be a little unconventional.
In what follows, we shall be considering, in the notation of the text, a flux line liquid in a superconducting sample of 
thickness $L$ along the $\hat{\bf z}$ axis.
We will call $\Omega$ the area of the sample in the $(x,y)$ plane.
The self energy per unit length of a single flux line will be denoted by $\varepsilon_1$,
and is given by $\varepsilon_1=\varepsilon_0\ln(\lambda_{ab}/\xi_{ab})$, where 
$\lambda_{ab}$ and $\xi_{ab}$ are the London penetration depth and the coherence length in the basal $(ab)$ plane. 
The magnitude of the magnetic induction inside the sample will be denoted by $B$, and it
is related to the flux line density $\rho=N/\Omega$ by $B=\rho\phi_0$.

To start with, we make the observation that 
the chemical potential per unit length of a system of $N$ flux lines  
can be easily obtained using the thermodynamic identity
(the chemical potential here should not be confused with the mass of the internal modes of flux lines)
\begin{eqnarray}
\mu = \frac{1}{L}\,\frac{\partial F}{\partial N} \label{def-mu-thermo}
\end{eqnarray}
where $F$ is the Helmholtz free energy of the flux line system. Using the fact that the free energy per unit
volume $f=F/(\Omega L)$ satisfies the following thermodynamic equality
\begin{eqnarray}
\frac{\partial f}{\partial B} = \frac{H}{4\pi}
\end{eqnarray}
and remembering that $B=n\phi_0=N\phi_0/\Omega$, we readily obtain
\begin{eqnarray}
\mu = \frac{H\phi_0}{4\pi} \label{correct-mu}
\end{eqnarray}
Instead of the Helmholtz free energy $F$, Nelson and Seung\cite{Nelson-Seung} consider the Gibbs free energy 
$G = F - V(BH/4\pi)$, which for a system of flux lines is given by~: 
\begin{eqnarray}
G  & = & \sum_{i=1}^N\varepsilon_1\int dz\,\Big( 1 + (d{\bf r}_i/dz)^2\Big)^{1/2} +
\nonumber\\
& + &\frac{1}{2}\sum_{i\neq j}\int dz \;V[{\bf r}_i(z) - {\bf r}_j(z)] - V\;\frac{BH}{4\pi}
\end{eqnarray}
where $V({\bf r})=2\varepsilon_0K_0(r/\lambda_{ab})$ is the interaction potential between flux line 
elements at height $z$. 
Upon expanding the square root on the {\em rhs} of this last equation, and readjusting the tilt modulus 
$\varepsilon_1\to\kappa=\varepsilon^2\varepsilon_1$ to take anisotropy into account, (and using 
the fact that the volume $V=\Omega L$ and $B=N\phi_0/\Omega$), they obtain~:
\begin{eqnarray}
G  & = & \Big(\varepsilon_1-\frac{H\phi_0}{4\pi}\Big)\,NL +
\frac{1}{2}\,\kappa\sum_{i=1}^N\int dz\, (d{\bf r}_i/dz)^2 +
\nonumber\\
& + &\frac{1}{2}\sum_{i\neq j}\int dz \;V[{\bf r}_i(z) - {\bf r}_j(z)] 
\end{eqnarray}
They then identify the term proportional to $N$ in this last expression, i.e.
$\big(\varepsilon_1- H\phi_0/4\pi\big)NL$ with the ``usual'' chemical potential term $-\mu N$ and obtain in this 
way the chemical potential per unit length as 
\begin{eqnarray}
\mu = (H\phi_0/4\pi) -\varepsilon_1
\label{wrong-mu}
\end{eqnarray}
which does not coincide with expression (\ref{correct-mu}).

\medskip

At the origin of this disagreement is a confusion in terminology that is worth clarifying. When we use terms such 
as ``chemical potential'' in vortex physics, we are borrowing a terminology 
whose most natural realm is the theory of gases and
liquids. There, one defines 
the Helmholtz free energy $F$ by $F_{liq} = -{ }T\ln Z$, where $Z$ is the canonical
partition function of the system (we use the subscript $liq$ to distinguish quantities pertaining to liquid state 
theory from quantities in vortex physics bearing the same name). 
$F_{liq}$ contains information about the interactions among the particles in 
the system and possibly, as is the case for vortices, about their self-energies.
One also defines a Gibbs free energy $G_{liq}$ by $G_{liq}=\mu N$, where 
$\mu=(\partial F_{liq}/\partial N)$ is the chemical potential. In the grand-canonical ensemble, 
the grand canonical potential ${\cal Q}_{gr}=-{ }T\ln{\cal Z}_{gr}$ can be written in terms of $F_{liq}$ and 
$G_{liq}$ as ${\cal Q}_{gr} = F_{liq} - G_{liq} = F_{liq} - \mu N$. 
For homogeneous systems, the equilibrium density of particles $\rho$ 
is obtained by minimizing the quantity $({\cal Q}_{gr}/V) = f -\mu\rho$, using the equation 
\begin{eqnarray}
\frac{\partial{\cal Q}_{gr}}{\partial\rho} = 0 \label{min-Omega-homog} 
\end{eqnarray}
or if a one body 
external potential is present, using a functional derivative
\begin{eqnarray}
\frac{\delta{\cal Q}_{gr}}{\delta\rho({\bf r})} =0  \label{min-Omega-inhom}
\end{eqnarray}
This last equality is in fact the expression of the second Hohenberg-Kohn-Mermin theorem
\cite{Hohenberg-Kohn,Mermin} which is the basis of density functional theories of classical liquids.
 
\medskip

In the case of flux line systems, what we call Gibbs free
energy $G = F - V(BH/4\pi)$ corresponds to the grand
canonical potential ${\cal Q}_{gr}$ of liquid state theory, as can easily be seen  
by comparing the definition of $G$ for vortices, $G=F-L(\phi_0H/4\pi)N$, to
${\cal Q}_{gr} = F -\mu{N}$, and remembering that the
equilibrium density of flux lines $\rho=B/\phi_0$ is obtained in a way reminiscent of equation 
(\ref{min-Omega-homog}), i.e.
\begin{eqnarray}
\frac{\partial G}{\partial\rho}=\frac{\partial G}{\partial B} = 0
\end{eqnarray}
The fact that $G$ for magnetic systems corresponds to the grand canonical potential has been recognized a long
time ago by de Gennes \cite{deGennes}, who obtained the pressure in the flux line system as 
$P=-(\partial G/\partial\Omega)$, which is reminiscent of
thermodynamic relation ${\cal Q}_{gc}= - PV$ (with $V$ the volume of the system). 
In view of the fact that the Helmholtz free energy $F$ has the same meaning in liquid state theory and for
magnetic systems, one is led to identify $V(BH/4\pi)$ with the Gibbs free energy $\mu N$ of liquid state
theory, and this identification leads to the result (\ref{correct-mu}) for the chemical potential per unit
length of the flux line system (see also the table below, which contains a summary of the correspondence between 
flux line and liquid quantities)

\medskip

We now can understand the problem with the derivation of the chemical potential $\mu$ in reference
\cite{Nelson-Seung}.The problem with this derivation is
that the term $\varepsilon_1 NL$ belongs to the Helmholtz free energy $F$ while the true chemical
potential should be identified only form the $G_{liq}$ part of the grand canonical potential, i.e. from
$V(BH/4\pi)=L(\phi_0H/4\pi)N$.
This improper identification of the chemical potential $\mu$ is, as far as the results of 
refs.\cite{Nelson-Seung,Nelson-LeDoussal,Tauber-Nelson} are concerned, only a minor, 
``cosmetic'' point. However, in order to 
obtain meaningful results in a grand-canonical formulation (using an approach similar to ours in (I) and in the 
present work), it is very important that the correct expression of $\mu$, 
equation (\ref{correct-mu}), is properly identified.

\medskip

\begin{center}
\begin{tabular}{ c  c  c }
\hline\hline
{ } & \hspace{1.5cm}  & { }\\
Flux lines  &      {     }               &  Classical Fluids \\
{ } &  { }  & { } \\
\hline \\
{Helmholtz free}  &    { }           &         {Helmholtz free} \\ 
{energy}          &    { }           &         {energy}  \\
{ } & {}  & { } \\
$F = -{ }T\ln Z$     & { }  &   $F = -{ }T\ln Z$ \\
{ } & { }  & { } \\
\hline \\
Gibbs free energy  & { }  & Grand canonical \\
                   &  { }  & potential \\
{ } & { }  & { } \\
$G = F - L(\phi_0H/4\pi)N$ & { }  &  ${\cal Q}_{gr}= F - {\mu}N$  \\
{ } &           { }        &  { } \\
$(\partial G/\partial B) = 0$   &  { }  & $(\partial{\cal Q}_{gr}/\partial\rho) = 0$ \\
{ }          &        { }         &           { } \\
\hline\\
chemical potential  & { }    & chemical potential \\
per unit length    & { } &          \\
{ } & { }  & { } \\
$\mu = (\partial f/\partial\rho)/L$ & { } & $\mu = (\partial f/\partial\rho)$ \\
{ } & { }  & { } \\
$ = (\phi_0H/4\pi)$   & { }  & { }  \\
{ } & { }  & { } \\
\hline\hline
\end{tabular}
\end{center}

\smallskip

{\em Table 1.} Summary of correspondence between vortex and classical fluids quantities.

\end{multicols}

\begin{references}
\bibitem{Remark}
Note that, had we chosen to use the most general form of the Hamiltonian of interacting flux lines in anisotropic 
superconductors \cite{Barford,Sardella,Sardella,Brandt}, we would have had to use an effective Hamiltonian for 
the internal
modes $H_u  =  \frac{1}{2}\,\sum_{i,j}\sum_{n\neq 0}\;[G^{-1}(q_n)]_{ij}^{\alpha\beta}
u_{i\alpha}(q_n) u_{j\beta}(-q_n)$
with an elastic propagator $[G^{-1}(q_n)]_{ij}^{\alpha\beta}$ which is non-diagonal in the spatial indices
$\alpha,\beta$.
\end{references}
\end{document}